# Magnetoelectric topology: the rope weaving in parameter space

Ying Zhou, Ziwen Wang, Fan Wang, Haoshen Ye, Shuai Dong[†]

*Key Laboratory of Quantum Materials and Devices of Ministry of Education, School of Physics, Southeast University, Nanjing 211189, China*

Topology, as a mathematical concept, has been introduced into condensed matter physics since the discovery of quantum Hall effect, which characterizes new physical scenario beyond the Landau theory. The topologically protected physical quantities, such as the dissipationless quantum transport of edge/surface states as well as magnetic/dipole quasi-particles like skyrmions/bimerons, have attracted great research enthusiasms in the past decades. In recent years, another kind of topology in condensed matter was revealed in the magnetoelectric parameter space of multiferroics, which deepens our understanding of magnetoelectric physics. This topical review summarizes recent advances in this area, involving three type-II multiferroics. With magnetism-induced ferroelectricity, topological behaviors can be manifested during the magnetoelectric switching processes driven by magnetic/electric fields, such as Roman-surface/Riemann-surface magnetoelectricity and magnetic crankshaft. These exotic topological magnetoelectric behaviors may be helpful to pursuit energy-efficient and precise-control devices for spintronics and quantum computing.

## 1. Introduction

The study of materials and phenomena with nontrivial topological structures/behaviors is both scientifically important and technologically promising. In a mathematical context, topology characterizes properties of geometric objects that remain invariant under continuous deformation, such as stretching or bending, but not tearing or gluing. A classic illustration is that a donut and a coffee cup are topologically equivalent because each has one hole (the donut's center hole and the coffee cup's handle) and one can be imagined as being molded into the other through continuous deformation [1-4].

In early studies of condensed matter, the concept of topology was mainly applied to the electronic structure in momentum space. In this context, systems such as various topological insulators, semimetals, and superconductors display nontrivial band topology, leading to novel phenomena like the quantum anomalous Hall effect [5-10]. These phases are defined by the topological invariants of bands, such as the Chern number and the $Z_2$ index, and have been a hot spot of condensed matter for more than two decades [11-13].

In addition, the concept of topology has also been introduced to describe some structures in real space, especially in magnetic and ferroelectric systems [14-17]. Topologically protected spin or polarization textures, such as skyrmions in $Cu_2OSeO_3$ and vortices of domains in hexagonal manganites, have drawn significant attentions for their stability against perturbations and their potential in spintronic and memory devices [18-22]. These textures are typically characterized by an integer winding number defined over a spherical manifold.

---
[†] Corresponding author. E-mail: sdong@seu.edu.cn

These manifestations of topology in momentum space and real space form different branches of topological entities in matter, as summarized in Fig. 1.

The role of topology in condensed matter systems can go beyond the momentum space and real space. From a broader mathematical view, any nontrivial structure or topological invariant in a given physical space can be seen as a manifestation of topology. Recent studies have extended this idea to more general spaces, particularly the magnetoelectric (ME) phase space, where unusual topological behaviors have been observed in complex multiferroic materials (Fig. 1) [23-27]. This space is parameterized by the components of the magnetization and polarization, as well as their corresponding external fields. The topology arises when the coupled evolution of magnetization and polarization under cyclic fields traces out non-trivial paths in this space, which can be mathematically described as weaving on complex manifolds. This formalism is particularly powerful in type-II multiferroics, where ferroelectricity is directly induced by specific magnetic orders, such as some non-collinear magnetic structures, leading to intrinsically strong ME coupling [28-30].

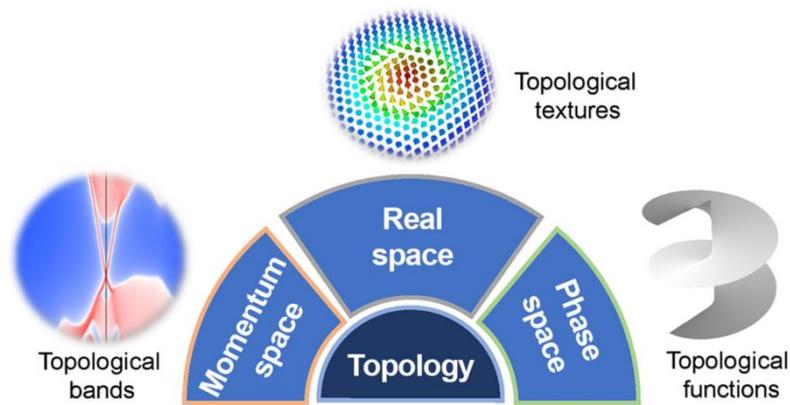

Fig. 1. Classification of topological phenomena in matter, encompassing topological band structures in momentum space, topological textures in real space, and topological functions in parameter phase space. Reprinted with permission from Ref. [26]. Copyright 2024 by American Physical Society.

The novelty of magnetoelectric topology lies in its ability to redefine our understanding of magnetoelectric coupling and mutual control. It transcends the conventional linear magnetoelectric effect, which describes a simple proportional relationship. Instead, it reveals that the functional relationship between magnetization and polarization can itself possess a robust, topologically protected structure. This guarantees the quantized and dissipationless switching of states during field cycles, offering unprecedented precision and reliability for device engineering.

In this topical review, we will systematically introduce some recent progresses in topological magnetoelectricity. Section 2 discusses the discovery of a non-orientable Roman surface governing polarization vectors in the cubic perovskite TbMn$_3$Cr$_4$O$_{12}$, where Néel vector trajectories on a sphere map to path-dependent magnetoelectric responses. Section 3 describes the realization of double-leaf Riemann surface magnetoelectricity in the GdI$_2$ monolayer, where the protected topology allows a precise 180° spin reversal with a single cycle of electric field, resulting in robust and dissipationless converse magnetoelectric functionality. Section 4 focuses on topologically protected magnetoelectric switching in GdMn$_2$O$_5$, showing how magnetic field cycles drive half spins to rotate continuously with a quantized winding number, while the rest half spins only toggles.

## 2. Realization of topological magnetoelectricity
### 2.1. Topological Roman surface in cubic lattices.

The quadruple perovskites can host the so-called type-II multiferroicity. The first example was CaMn$_7$O$_{12}$, which owns a noncollinear spin order below 90 K [31, 32]. The magnetism-induced ferroelectric polarization can be understood as the combined effect of Dzyaloshinskii-Moriya interaction and exchange striction [33]. Later, the type-II multiferroicity was observed in cubic LaMn$_3$Cr$_4$O$_{12}$ [34], as well as TbMn$_3$Cr$_4$O$_{12}$ (TMCO) [19], which opens the door of topological Roman surface magnetoelectricity.

TMCO crystallizes in the highly symmetric cubic perovskite structure with space group $Im\bar{3}$. This A-site ordered variant adopts the general formula AA'$_3$B$_4$O$_{12}$ [35, 36], where Tb and Mn ions occupy the 2$a$ (0, 0, 0) and 6$b$ (0, 0.5, 0.5) Wyckoff positions, respectively, in a 1:3 ordered arrangement. The Cr ions are located at the 8$c$ (0.25, 0.25, 0.25) sites, while oxygen ions reside at the 24$g$ ($x$, $y$, 0) sites [37, 38]. Below its Néel temperature $T_{N2} \approx 36$ K, the Mn$^{3+}$ and Cr$^{3+}$ sublattices each forms a collinear G-type antiferromagnetic structure, respectively [34] (Fig. 2(a)). Such a magnetic order breaks the spatial inversion symmetry and induces an electric polarization along the [111] direction, mediated by the anisotropic symmetric exchange interactions [39].

Remarkably, the cubic symmetry constrains the polarization vector *P*--parametrized by the orientation of the total Néel vector (represented by **S**$_{Cr}$) on a unit sphere **S**$^2$--to lie on a so-called Roman surface, a non-orientable two-dimensional manifold [23] (Figs. 2(b-c)). The mathematical form of this Roman surface of induced polarization is given by [23]:

$$P_x^2 P_y^2 + P_x^2 P_z^2 + P_y^2 P_z^2 - \delta P_x P_y P_z = 0, \qquad (1)$$

where $\delta$ is a material-dependent parameter, and $P_{x/y/z}$ are components of polarization along the cubic $x/y/z$ axes, respectively.

The Roman surface is non-orientable with one triple point, six pinch points, and three double lines, and shows different winding numbers depending on the path (Fig. 2(c)). Such nontrivial topology can lead to some novel magnetoelectric behavior. For example, a 180° rotation of the **S**$_{Cr}$ Néel vector makes the polarization vector back to itself. Paths that go around the double lines have a winding number $Q$=1 and show the $2\pi$ periodicity, while paths that avoid singularities have a winding number $Q$=0.

For polycrystalline TbMn$_3$Cr$_4$O$_{12}$, this topological behavior is experimentally accessible by rotating magnetic fields, which steers the Néel vector trajectories on **S**$^2$ and induces a corresponding evolution of **P** on **RP**$^2$ (**RP**$^2$, real projective plane) (Fig. 2(d)). In low-field regimes (e.g. $B$~2 Tesla), **S**$_{Cr}$ undergoes small tilts, generating trivial loops that result in double-angle sinusoidal modulation of polarization without changing the sign (Fig. 2(e)). When the magnetic field exceeds a critical value $B_c \approx 3$ Tesla, spin-flop transitions lead to nontrivial trajectories with $Q$=1, resulting in sinusoidal polarization oscillations with sign reversal (Fig. 3(f)). Such behavior is also observed in the isostructural compound LaMn$_3$Cr$_4$O$_{12}$, confirming the universality of this magnetoelectric topology in the

$A$Mn$_3$Cr$_4$O$_{12}$ family [23].

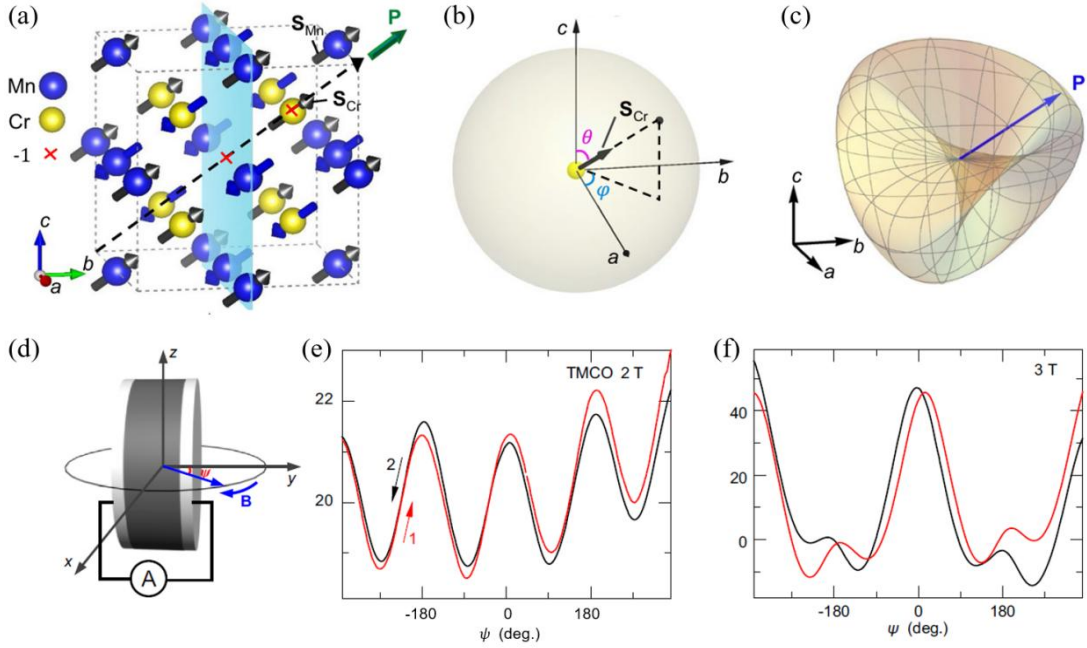

Fig. 2. Topological Roman surface magnetoelectricity of TbMn$_3$Cr$_4$O$_{12}$. (a) Crystal structure (Tb and O ions omitted for clarity) with key symmetry elements (inversion centers $\bar{1}$, mirror plane, three-fold axis). The antiferromagnetic spin alignment of Mn and Cr sublattices (Néel vectors $\mathbf{S}_{Mn}$ and $\mathbf{S}_{Cr}$) is also shown. At the ground state, both $\mathbf{S}_{Cr}$ and the induced polarization $\mathbf{P}$ are along the [111] axis. (b) Orientation of $\mathbf{S}_{Cr}$, characterized by the spherical coordinates ($\mu_{Cr}$, $\theta$, $\varphi$). (c) The resulting polarization $\mathbf{P}$ forms a Roman surface upon the rotation of $\mathbf{S}_{Cr}$. (d) Experimental setup for measuring the magnetoelectric current while rotating $\mathbf{B}$ within the $xy$-plane, where $\psi$ is the angle between $\mathbf{B}$ and the $y$-axis. (e-f) The polarization $\mathbf{P}$ at 10 K of TMCO, obtained by integrating the magnetoelectric current, as a function of $\psi$ at $B$ = 2 T and 3 T. The red and black arrows denote the direction of the angle sweep. Reprinted with permission from Ref. [23]. Copyright (2022) by Authors.

Despite these experimental evidences, the nature of Roman surface magnetoelectric topology in $A$Mn$_3$Cr$_4$O$_{12}$ compounds has also been further confirmed by first-principles calculations [24]. In particular, the orientations of spins and polarizations can be deterministic in calculations, while they are not clear in aforementioned polycrystalline experiments. Wang *et al.* simulated the polarization evolution under continuous spin rotations while preserving the cubic $Im\bar{3}$ symmetry. Figure 3 illustrates the evolution of the magnetically induced dipole moment under a set of distinct closed spin-rotation paths. For four representative rotation loops (Figs. 3(a-d)), the three dipole components exhibit well-defined periodic modulations with respect to spin orientations (Figs. 3(e-h)), generating a variety of three-dimensional trajectories, including elliptical, saddle-like, vertical-line, and trefoil forms (Figs. 3(i-l)). The specific geometry of each trajectory, as well as its projection onto the $xy$ plane, is determined by the chosen rotation path. In all cases, the dipole vectors strictly conform to the mathematical constraint of a Roman surface.

To answer the question whether the applied magnetic field can rotate these two sublattices synchronously or just each sublattice independently, Wang *et al.* calculated the

DFT energy of the spin rotation of an individual sublattice with the other sublattice fixed to the [111] orientation, as shown in Figs. 4(a-b). The energy reaches the lowest one when the Cr/Mn sublattices are parallel or antiparallel (i.e. the ground state) but reaches ~24 meV/u.c. when the spins of Cr/Mn sublattices are perpendicular to each other. For comparison, the energy fluctuations associated with the synchronous rotation of two spin sublattices are only in the order of 0.1-0.4 μeV/u.c., significantly lower than the independent rotation of each sublattice. Thus, the energy barrier which locks the Néel vectors $S_{Cr}$ and $S_{Mn}$ to be parallel or antiparallel in the region of low energy excitation, leads to the synchronous rotation under magnetic fields.

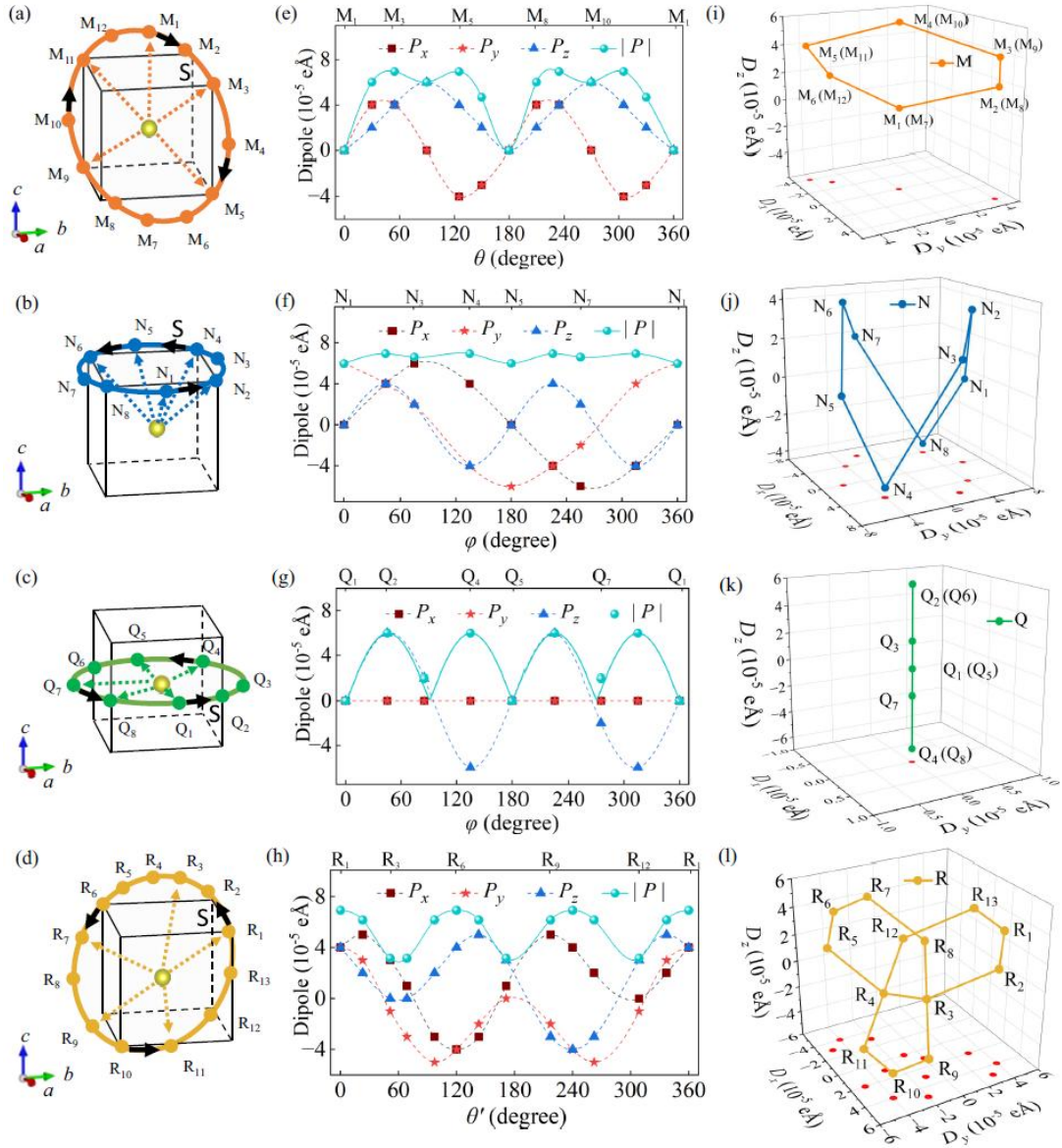

Fig. 3. Induced electric dipoles resulting from distinct closed paths of spin rotation, calculated using the rigid high-symmetric crystal structure to prevent artifacts from lattice relaxation. (a-d) Four characteristic rotation paths: (a) the longitudinal ($\theta$ from 0° to 180° at $\phi$=45°/225°); (b, c) the latitudinal ($\varphi$ from 0° to 360° at $\theta$=54.7° and 90°); (d) a path within the (1$\bar{1}$1) plane. (e-h) The corresponding evolution of the dipole vector components. (i-l) Three-dimensional trajectories of the dipole vector for sampling points M, N, Q, R, with red dots indicating their projections onto the *xy*-plane. Notably, the paths in (i) and (k) yield identical

trajectories over both half-periods. All trajectories are constrained to the topological Roman surface. Reprinted with permission from Ref. [24]. Copyright 2023 by American Physical Society.

In addition to confirming the topology, the calculations clarify its physical origin. The spin-induced polarization arises as a second-order effect of spin-orbit coupling (SOC), scaling with the SOC coefficient $\lambda^2$ [40], as shown in Fig. 4(c). Based on this insight, it is possible to enhance the polarization by substituting B-site $Cr^{3+}$ with heavier ions. For instance, $BaMn_3Re_4O_{12}$ has the same magnetic structure but a much larger polarization (~810 μC/m²), while keeping the Roman surface topology. The $Re^{4+}$ ion can not only enhance the effective SOC effect of B-site, but also active the effective SOC in the A'-site by changing the valence of Mn from 3+ (almost SOC quenched) to 2+. This attempt may offer a route toward realizing topological multiferroicity with stronger and detectable ME effects.

Unlike conventional real-space topological textures such as skyrmions or vortices [41, 42], the Roman surface here governs a dynamical mapping between spin and polarization vectors in the magnetoelectric parameter space. These findings highlight the potential of high-symmetry perovskites as ideal platforms for topological control of multiferroic properties via external magnetic fields.

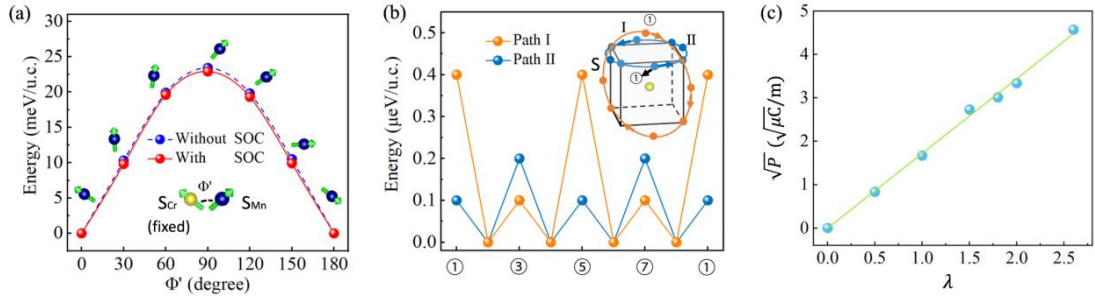

FIG. 4. (a) Energy barrier for $S_{Mn}$ rotation with fixed $S_{Cr}$. The identical profiles with/without SOC imply SOC plays a negligible role. This barrier sets the theoretical limit for switching between +**P** (0°) and -**P** (180°) states. (b) Energy change during synchronous rotation of both spin sublattices. (c) The square root of the induced polarization scales linearly with the SOC strength. Reprinted with permission from Ref. [24]. Copyright 2023 by American Physical Society.

## 2.2. Double-leaf Riemann surface topological converse magnetoelectricity in monolayer $GdI_2$

Besides the Roman surface, other exotic manifolds with nontrivial topology can also be achieved in the magnetoelectric parameter space. Here an example based on $GdI_2$ monolayer will be introduced, in which the magnetoelectric behavior forms a double-leaf Riemann surface.

The $GdI_2$ monolayer crystallizes in a noncentrosymmetric and nonpolar structure (space group $P\bar{6}m2$, No. 187), as shown in Fig. 5(a). It exhibits low exfoliation energy (0.26 J/m²), suggesting feasible mechanical isolation from its van der Waals bulk [43]. As shown in Fig. 5(b), Each $Gd^{2+}$ ion occupies a distorted trigonal prismatic coordination site, featuring a high-spin $4f^75d^1$ configuration with a magnetic moment of 8 $\mu_B$ [43]. The crystal field stabilizes the remaining 5d electron in a singlet state derived from $3z^2-r^2/x^2-y^2/xy$ orbitals [44], resulting in easy-plane anisotropy and a magnetocrystalline anisotropy energy (MAE) of ~0.7 meV/Gd (Fig. 5(c)). The monolayer is a ferromagnetic insulator with a Curie temperature near room temperature (~241 K) and a band gap of 0.6 eV (GGA+$U$) or 1.1 eV (HSE06) [39].

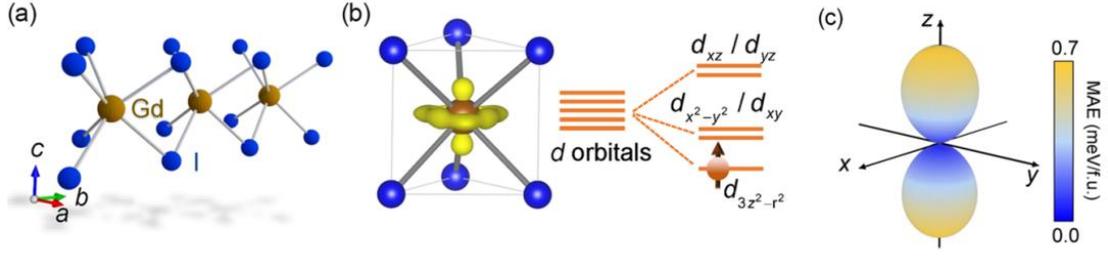

Fig. 5. Basic physical properties of GdI$_2$ monolayer. (a) Schematic of crystal structure. (b) Prism-type crystalline field splitting and the corresponding $5d^1$ electron occupancy, visualized by the calculated electron density. (c) The magnetic anisotropy energy as a function of spin orientation. Reprinted with permission from Ref. [26]. Copyright 2024 by American Physical Society.

Although its crystalline structure is nonpolar, the orientation of spin $\mathbf{S}_{Gd}$ may break the spatial inversion symmetry. When the spin $\mathbf{S}_{Gd}$ lies in its easy plane, the induced polarization ($P_x$, $P_y$, $P_z$) follows the form ($\cos2\varphi_s$, $-\sin2\varphi_s$, 0) ($\varphi_s$: the azimuth angle of spin) according to the mechanism of spin-dependent $p$-$d$ hybridization [26], in excellent agreement with the symmetry analysis and the first-principles calculations (Figs. 6(a-c)). This expression represents a polarization phase-doubling effect: when the in-plane $\mathbf{S}_{Gd}$ rotates, e.g. driven by a rotational in-plane magnetic field, the polarization vector will rotate in a twice angular speed of magnetic field. Its reciprocal process is that the magnetization undergoes only a half-cycle rotation when an in-plane rotational electric field is applied to drive a cycle rotation of polarization (Figs. 6(d-e)). This establishes a fixed 1:2 ratio for the winding numbers, namely odd winding numbers of electric component produce a robust 180° flip of magnetization, and even winding numbers restore the original magnetic state. The behavior is equivalent to a double-leaf Riemann surface, providing intrinsic topological protection for the magnetization reversal (Fig. 6(f)).

To further quantify this topological coupling, Figure 7 analyzes the system's dynamical response under both static and alternating electric fields. Under a constant electric field of 1 kV/cm along the -$x$ direction, the polarization reverses while the magnetization rotates by 90°. With this 1 kV/cm field, the characteristic switching time is in the order of 1 ns for GdI$_2$ monolayer, according to the Landau-Lifshitz-Gilbert (LLG) dynamic simulation (Figs. 7(a-c)). For alternating electric field (Fig. 7(d)), both $\mathbf{P}$ and $\mathbf{S}_{Gd}$ fully follow the oscillating electric field in the low-frequency (e.g. 50 MHz) regime, but they only oscillate with small amplitude around their middle positions in the high frequency (e.g. 500 MHz) regime.

Using a four-cross-electrode configuration to alternately apply electric field pulses (Fig. 7(e)), each pulse drives a 90º rotation of $\mathbf{P}$ and 45º rotation of $\mathbf{S}_{Gd}$. After eight pulses, a full rotation of $\mathbf{S}_{Gd}$ and two cycles of $\mathbf{P}$ are completed, as shown in Fig. 7(f).

The underlying mechanism of double-leaf Riemann surface topology is not limited to the GdI$_2$ monolayer only [26]. We found that similar topological behaviors can emerge in other materials with appropriate symmetry, particularly those lacking inversion symmetry and exhibiting strong spin-orbit coupling (SOC). By classifying materials based on symmetry, we determined that only systems with threefold rotational symmetry in the plane, such as certain two-dimensional ferromagnets, can exhibit this topological magnetoelectric behavior [45]. Materials like 2H-$MX_2$-type monolayers and $VSi_2N_4$ monolayers, which meet these symmetry conditions, are potential candidates for further study [26].

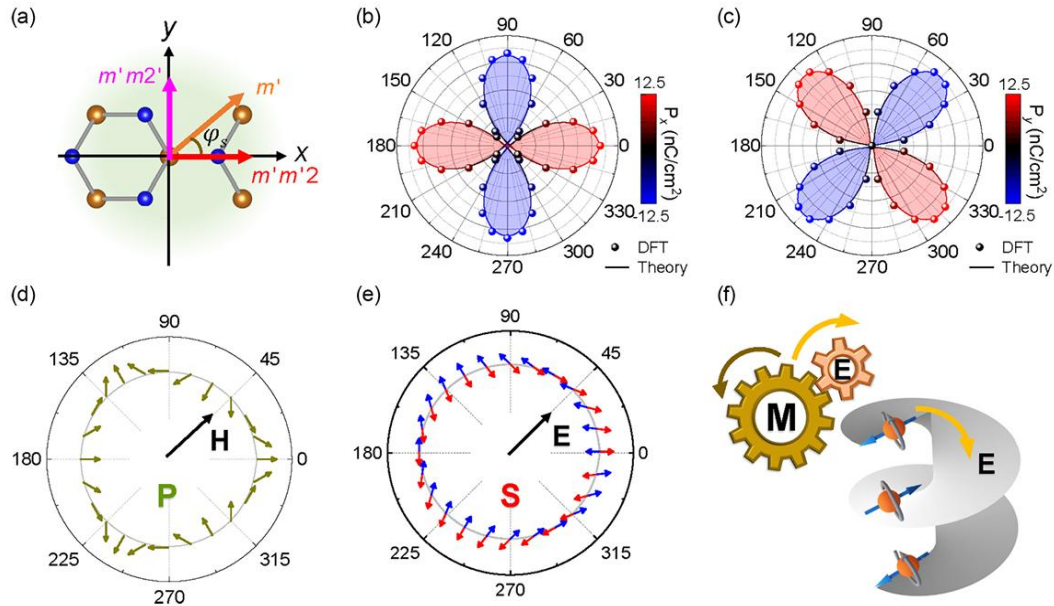

Fig. 6. Topological magnetoelectricity in $GdI_2$ monolayer, described by a double-leaf Riemann surface manifold. (a) Magnetic symmetry (MPG) as a function of Gd spin angle $\varphi_s$. (b, c) Density functional theory (DFT)-calculated in-plane polarization (dots) and theoretical fitting (curves) versus $\varphi_s$. (d) Polarization evolution under a rotating magnetic field. (e, f) The nonreciprocal response: a 360° rotation of electric field **E** rotates magnetization **M** by 180°, forming a 1:2 gear relation. This topology guarantees that the magnetization is flipped precisely by 180° for any odd winding number of **E**, and restored only by an even winding number. Reprinted with permission from Ref. [26]. Copyright 2024 by American Physical Society.

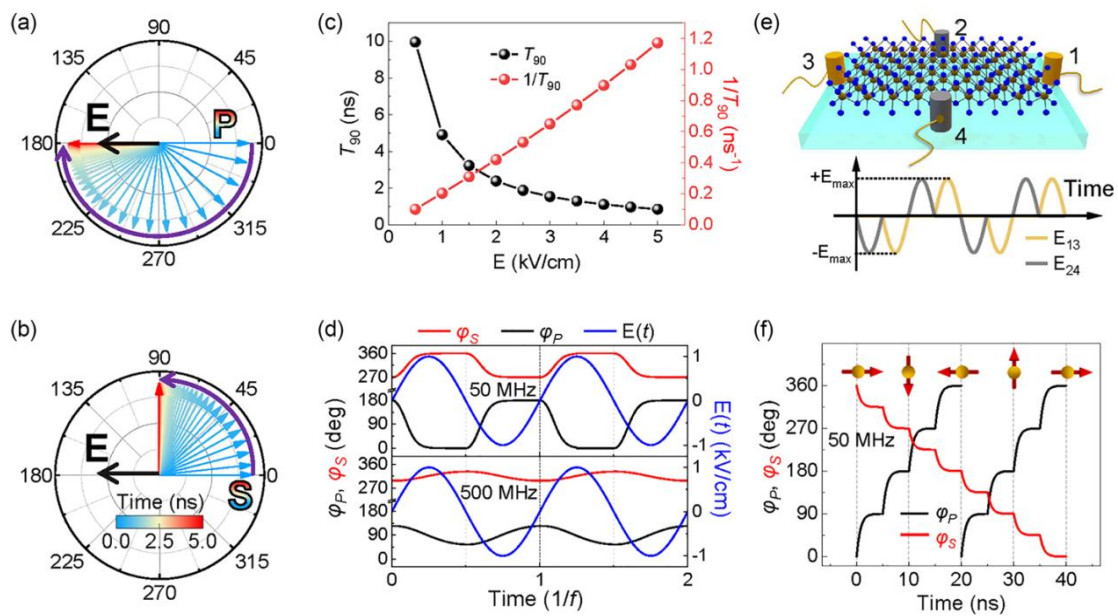

Fig. 7. Dynamic response of polarization **P** and Gd spin $S_{Gd}$ to applied electric fields. (a, b) Switching dynamics triggered by a 1 kV/cm static field. (c) Field dependence of the switching speed ($1/T_{90}$). (d) Oscillatory behavior of $\varphi_P$ and $\varphi_s$ under 50 MHz and 500 MHz alternating electric field. (e) Schematic diagram of four electrodes in perpendicular configuration and pulse sequence. (f) Demonstration of half-harmonic spin precession under a modulated 50 MHz pulse train. Initial conditions for **P** and $S_{Gd}$ are along $+x$; all alternating electric field maxima are 1 kV/cm. Reprinted with permission from Ref. [26]. Copyright 2024 by American Physical Society.

## 2.3. Topologically protected magnetic crankshaft in GdMn$_2$O$_5$

Another material known for its large magnetism-induced polarization is GdMn$_2$O$_5$. It is a member of the multiferroic $R$Mn$_2$O$_5$ family ($R$: rare earth element), crystallizing in an orthorhombic structure with the centrosymmetric space group *Pbam* (No. 55) [46-50]. Mn$^{4+}$ ions occupy the octahedral sites while Mn$^{3+}$ ions are pyramidally coordinated, together forming zigzag chains along the $a$-axis, as shown in Fig. 8. This geometry leads to competing antiferromagnetic interactions, resulting in strong magnetic frustration [51, 52].

Upon cooling, GdMn$_2$O$_5$ undergoes a series of magnetic phase transitions. It first enters an incommensurate antiferromagnetic state below $T_{N1} \approx 40$ K, which then locks into a commensurate state with a propagation vector of (1/2, 0, 0) below $T_{N2} \approx 33$ K. In this low-temperature commensurate phase, two distinct antiferromagnetic chains develop along the $a$-axis. The symmetric exchange striction between these chains breaks spatial inversion symmetry and induces a spontaneous ferroelectric polarization, primarily along the $b$-axis, following the relation $P \sim L_1 \cdot L_2$ (where $L_1$ and $L_2$ are the antiferromagnetic vectors of the two chains). This mechanism yields one of the largest magnetically induced polarizations known, reaching up to 3600 μC/m$^2$ [53-55]. The presence of competing interactions gives rise to a rich phase diagram with multiple ferroelectric phases and diverse magnetoelectric effects, making GdMn$_2$O$_5$ an exceptional material for studying strong magnetoelectric coupling.

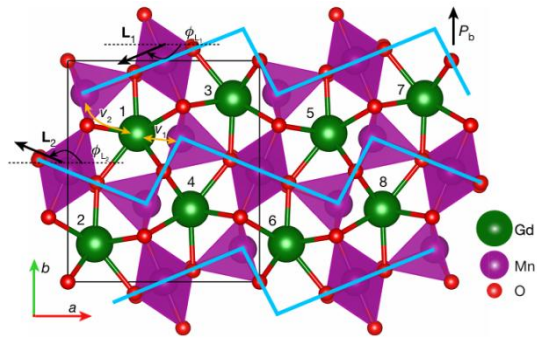

Fig. 8. Magnetic unit cell of GdMn$_2$O$_5$, showing Mn zigzag chains with intrachain antiferromagnetic order ($L_1$ and $L_2$ at angles $\phi_1$, $\phi_2$). Gd ions occupy pentagonal sites between chains, with oxygen ions mediating the superexchange interactions. Black rectangle: the structural unit cell. Reprinted with permission from Ref. [25]. Copyright 2022 by Springer Nature.

Remarkably, Artyukhin *et al.* revealed that the electric polarization in multiferroic

GdMn$_2$O$_5$ can be nontrivially manipulated and reversed by a magnetic field [25]. As shown in Fig. 9, the evolution of polarization strongly depends on the orientation of magnetic field and the temperature. When the field is far from the "magic angle" direction, conventional two-state switching is observed, with the polarization reversing only once per field cycle (Figs. 9(a) and 9(c)). However, at tilt angles close to ±10°, the system enters the four-state switching regime, where two distinct low- and high-field states are visited in sequence, and the polarization reverses twice per cycle (Figs. 9(b) and 9(d)).

This behaviour is symmetric for positive and negative tilt angles and is most pronounced at low temperatures. At higher temperatures, the hysteresis shrinks to a small step with conventional two-state behaviour (Fig. 9(e)), while the four-state loop gradually appears below about 5 K (Fig. 9(f)). These observations highlight that the magic-angle condition and low temperature are essential for realizing the topologically protected four-state switching in this material.

The microscopic mechanism is further illustrated schematically in Figs. 9(g-i), the specific switching pathway depends on the tilt angle of magnetic field: at large tilt angles, both Néel vectors toggle between two orientations; at small tilt angles near the crystallographic *a* axis, they toggle simultaneously; and only near the "magic angles" (~ ±10º), a four-state cycle emerges in which one Néel vector rotates continuously while the other toggles. From the aspect of topology, the two magnetic field cycles give rise to a nontrivial winding number $Q=\pm 1$ for the one spin chain.

Although above work established that the four-state switching, it occurred only under a precisely aligned magnetic field (e.g., the so-called "magic angle") at low temperatures. A recent work by Lu *et al*. demonstrated that this behavior can be realized over a broader range of field orientations and higher temperatures [27]. A key development is to use an external electric field to assist the topological switching process. As illustrated in Fig. 10(a), a positive electric field during the magnetic-field sweep produces a reversible two-state loop, whereas a negative electric field drives the system along an irreversible path that inverts the polarization. Reversing the electric field in the subsequent cycle (Fig. 10(b)) enables the system to traverse the complete four-state loop beyond the "magic angle", and return to its initial configuration after two magnetic cycles. This assistant electric field not only expands the stability range of topologically protected switching to higher temperatures up to the ferroelectric Curie temperature of ~33 K (Fig. 10(c)), but also relaxes the requirement for precise magnetic-field alignment (Figs. 10(d-g)). Thereby, it offers a more versatile and practically viable route to exploit this phenomenon.

Theoretical simulations that incorporate both magnetic and electric fields reveal the underlying mechanism (Fig. 11). During the magnetic sweeping cycles, the electric field breaks the degeneracy between states connected by inversion symmetry (e.g., the states 1/3 and 2/4). Thereby, the original symmetric energy landscape is biased to be asymmetric, which deterministically steers the spin rotation along a specific switching path during the magnetic field cycle. The critical role of the spin-flop transition, responsible for the unidirectional spin rotation, was also confirmed in the simulations.

This is a significant improvement by delicately tuning the energy landscape for topological switching. By violating the strict conditions of specific magnetic field orientation and low temperatures, it is hopeful to expect more similar four-state switching behavior in other members of $R$Mn$_2$O$_5$ family, as well as other magnetic ferroelectrics.

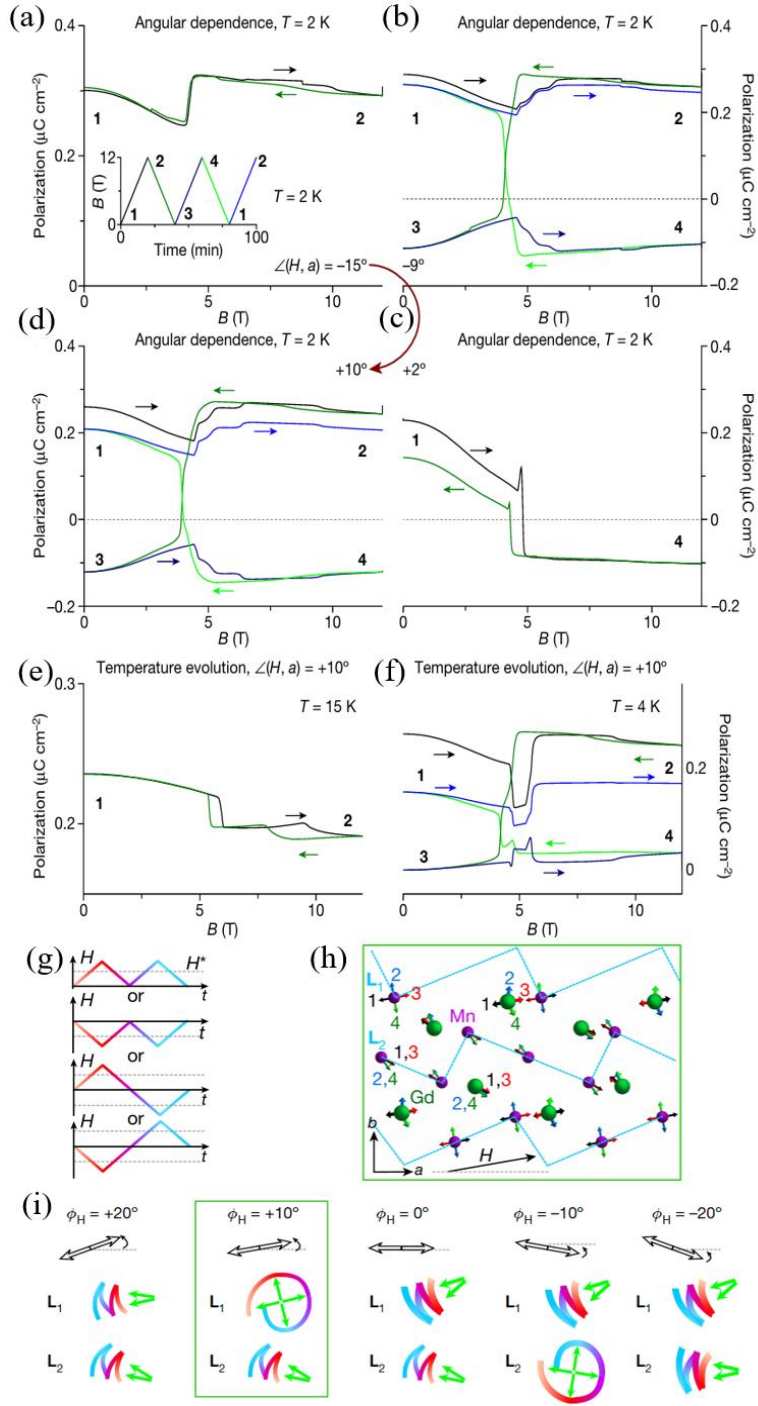

Fig. 9. Topologically magnetoelectric switching in GdMn$_2$O$_5$. (a-f) Angle- and temperature-dependent polarization hysteresis. (a-d) At 2 K, the **P-B** loop transitions from a standard two-state switch to a four-state loop as the magnetic field direction approaches the "magic angles" of ±10°. (e) At high temperatures, only a simple hysteretic step is observed. (f) The four-state switching behavior develops below $T$ = 5 K. (g) Time dependence of the magnetic field. The field is tilted by the angle $\phi_H$ away from the $a$ axis. Colour coding is used to indicate the synchronous evolution of the Néel order parameters, **L**$_1$ and **L**$_2$ in (i). For a fixed $\phi_H$, each of the shown magnetic-field ramp protocols (positive or negative field ramps) leads to the same result. (h) Spin configurations 1–4 in the magnetic unit cell, visited in the switching

process at $\phi_H = 10°$. (i) Comparison of topological trivial and nontrivial switching of antiferromagnetic chains. A color code indicates the concurrent evolution of the Néel order parameters, $L_1$ and $L_2$, under the magnetic sweep. Only when the magnetic field is along the the "magic angles", one Néel vector rotates continuously while the other toggles. Reprinted with permission from Ref. [25]. Copyright 2022 by Springer Nature.

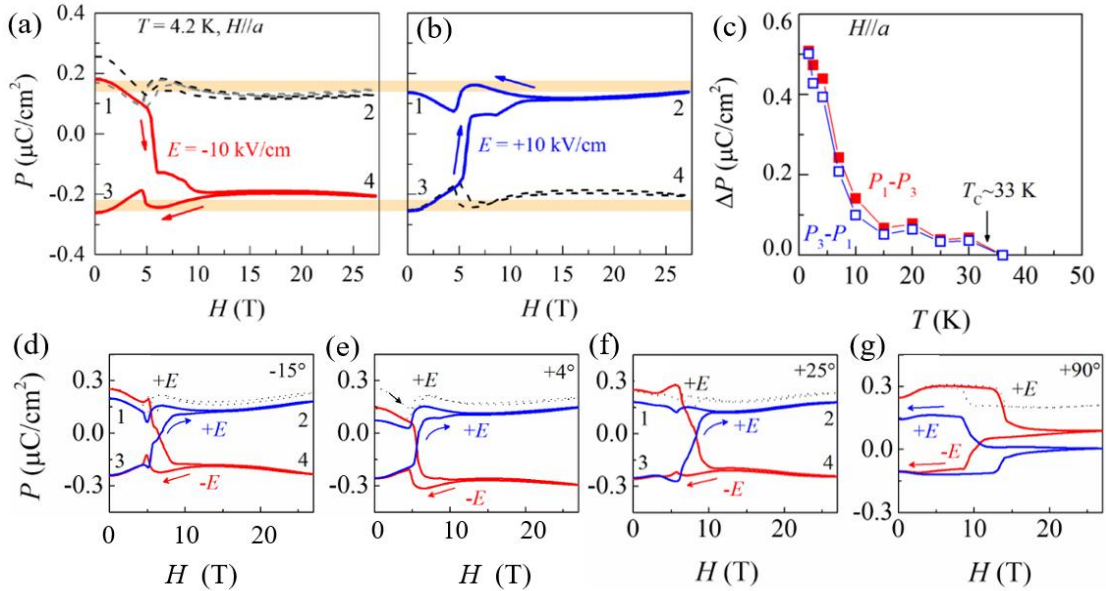

Fig. 10. Electric field assistant topological magnetoelectric switching in GdMn$_2$O$_5$. (a, b) Hysteresis loops at 4.2 K ($H//a$) showing electric-field-dependent switching among four states. (c) Temperature dependence of $\Delta P$ up to $T_C$. (d-g) Evolution of the $P$-$H$ loop shape with the in-plane angle $\theta$ of the magnetic field, measured under $E=\pm10$ kV/cm. Reprinted with permission from Ref. [27]. Copyright 2025 by American Physical Society.

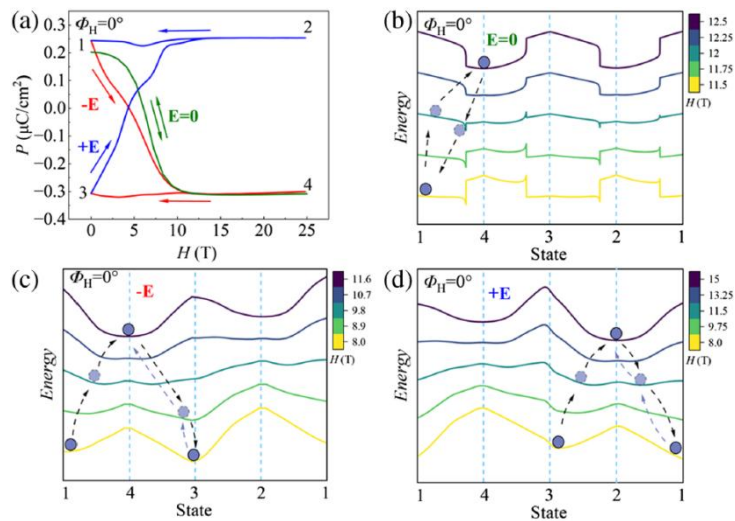

Fig. 11. Simulated magnetoelectric switching dynamics. (a) Simulated polarization ($P$) hysteresis cycle for $H//a$, comparing scenarios in the absence and presence of an applied $E$. (b-d) The corresponding free-energy landscapes for $H//a$. (b) Without $E$, the system exhibits a reversible transition path during the magnetic cycle. (c) Under a negative $E$, the $-P$ states (3, 4) become slightly energetically favored. (d) Under a positive $E$, the $+P$ states (1, 2) are slightly preferred. All transition paths are determined via the steepest descent method on the free-energy surface, rather than being predefined by the end states. Reprinted with permission from Ref. [27]. Copyright 2025 by American Physical Society.

## 3. Conclusion and perspectives

We have systematically reviewed the research progress of magnetoelectric topology in the type-II multiferroics. It is a conceptually new branch of topology-related physics in condensed matter. In these magnetic ferroelectrics, the magnetoelectric switching trace is nontrivial under periodic driving fields, forming topological nontrivial phenomena. A key characteristic is that these systems do not return to its original state after one full magnetic/electric field cycle, while a sequential cycle is required to reset its configuration. Mathematically, these behaviors can be described by Roman surface, Riemann surface, as well as winding numbers. However, the three prototypical materials discussed in our review ($TbMn_3Cr_4O_{12}$, $GdI_2$ monolayer, and $GdMn_2O_5$) currently exhibit distinct characteristics, as preliminarily summarized in Table 1. The field of magnetoelectric topology is still in its infancy stage, and the diversity of these pioneering examples highlights the rich phenomenology yet to be fully systematized.

Table 1. Classification of magnetoelectric topology.

| Systems | Crystalline symmetry | Order parameter | Topological invariants | Other candidates |
|---|---|---|---|---|
| $TbMn_3Cr_4O_{12}$ [23] | Centrosymmetric | Polarization | Winding number | $LaMn_3Cr_4O_{12}$ [23] $BaMn_3Re_4O_{12}$[24] |
| $GdI_2$ monolayer [26] | Non-centrosymmetric and non-polar | Magnetization/Polarization | Winding number | $VSe_2$/$VTe_2$ [56] |
| $GdMn_2O_5$ [25] | Centrosymmetric | Antiferromagnetic order | Winding number | - |

With precise and robust relationship between the electric and magnetic order parameters, the ability to manipulate topological magnetoelectric effects may offer opportunity for the development of quantum computing devices. It guarantees quantized and dissipationless (or low dissipative) switching of states, which is a cornerstone for developing low power and highly robust electronic devices. This unique property opens transformative opportunities in information storage.

As an emerging field, there remains many challenges to fully understand and utilize the magnetoelectric topology. The number of known materials exhibiting these effects remains very limited, with most being antiferromagnetic systems and faint polarizations. This material constraint hinders the efficient electrical control of magnetic states and practical implementation. To overcome these material constraints, future research must prioritize innovative materials design strategies aimed at achieving room-temperature multiferroics with robust topological ME effects. Promising avenues include: (i) exploring new material classes such as ferromagnetic and ferrimagnetic compounds where net magnetization facilitates detection and control; (ii) engineering heterostructures and interfaces to create emergent topological ME phenomena from the coupling between a strong ferromagnet and a ferroelectric; and (iii) employing high-throughput computational screening to identify promising candidates with desired symmetries and strong spin-lattice coupling. Beyond materials, a deeper theoretical understanding is required. A complete physical scenario that links microscopic spin dynamics, lattice distortions, and the emergent global topology in parameter space is essential. This will guide the intelligent design of materials and the prediction of new topological phenomena.

The experimental study of magnetoelectric topology is still in its early stages. Current characterization relies mainly on macroscopic magnetoelectric measurements, while direct microscopic observation is yet to be achieved. Promising techniques like the magnetoelectric probe developed by Wu *et al*. could potentially map antiferromagnetic domain walls and their evolution [57], offering a path forward. We look forward to future breakthroughs in microscopic characterization that will illuminate the local mechanisms of these topological effects.

In conclusion, magnetoelectric topology opens a vibrant and uncharted territory at the intersection of topology, multiferroicity, and spintronics. While challenges in materials discovery and theoretical understanding persist, the potential rewards are immense. Success in this direction will undoubtedly pave the way for a new era of topologically protected electronic functionalities.

**Acknowledgment**
This work was supported by National Natural Science Foundation of China (Grants Nos. 12325401, 12274069, 123B2053). We thank Profs. Yisheng Chai, Chenliang Lu, and Junting Zhang for their contributions to the original works reviewed here.